\documentclass[cameraready]{Interspeech}
\usepackage[utf8]{inputenc}
\usepackage{booktabs}   
\usepackage{multirow}   
\usepackage{array}      
\usepackage{url}
\newcommand{\cmark}{\ding{51}}
\newcommand{\xmark}{\ding{55}}
\usepackage{pifont}

\title{Entropy-Aware Domain-Routed Mixture-of-Experts Speech-LLM Framework: A Case Study of Multi-Domain Child-Adult ASR}

\author[affiliation={1}]{Mohan}{Shi}
\author[affiliation={1}]{Kaiyuan}{Zhang}
\author[affiliation={1}]{Zilai}{Wang}
\author[affiliation={1}]{Natarajan Balaji}{Shankar}
\author[affiliation={1}]{Eray}{Eren}
\author[affiliation={1}]{Abeer}{Alwan}

\address{
    $^1$ University of California, Los Angeles, USA
}

\email{ \{shimohan, kaiyuanzhang, zilaiwang2001, balaji1312, erayeren\}@g.ucla.edu, alwan@ee.ucla.edu}


\usepackage{comment}

\begin{document}

\maketitle

\begin{abstract}
    \vspace{-0.2cm}


    {While Speech Large Language Models (Speech-LLMs) have achieved strong performance on adult Automatic Speech Recognition (ASR), their effectiveness on child speech remains under-explored, and single models often struggle to handle diverse adult and child age groups simultaneously.}
    This paper proposes a Mixture-of-Experts (MoE) Speech-LLM for unified ASR across adult and child speech spanning diverse environments and age groups. The framework employs a Classifier-based Domain Router (C-DR) with a coarse-to-fine strategy and integrates both a Mixture-of-Projectors (MoP) and a Mixture-of-LoRAs (MoL) to model domain-specific variations. To address routing uncertainty near domain boundaries, an Entropy-Aware Routing (EAR) mechanism is introduced to dynamically incorporate a shared expert. Experiments on public child corpora demonstrate consistent improvements over baselines while preserving adult ASR performance. To our knowledge, this is the first work leveraging Speech-LLMs for unified, multi-domain ASR encompassing both children and adults.
\end{abstract}

\vspace{-0.2cm}
\section{Introduction}
Speech Large Language Models (Speech-LLMs) have demonstrated strong performance on a wide range of speech-related tasks~\cite{FathullahWLJSLG24,ma2024embarrassingly,ShiJXXZWSZY24,chen2025neural,TangYSC000M024,wang2026emotionthinker,shi2025trainshortinferlong}. 
A typical Speech-LLM for Automatic Speech Recognition (ASR) leverages a pre-trained speech encoder~\cite{BaevskiZMA20,HsuBTLSM21,ChenWCWLCLKYXWZ22}, a modality projector, and a  LLM~\cite{grattafiori2024llama,yang2025qwen3} fine-tuned with a Low-Rank Adapter (LoRA)~\cite{HuSWALWWC22}, and this framework has shown impressive ASR performance on adult speech under well-controlled conditions. 
However, child ASR~\cite{fan2022draft,fan2022towards,FanSA24,ying25_wocci} remains a longstanding challenge due to substantial acoustic and linguistic differences between child and adult speech, as well as the limited availability of large-scale, high-quality annotated child speech data. 
To date, no prior Speech-LLM work has demonstrated strong performance on child ASR benchmarks using public corpora, such as~\cite{PradhanCW24,ShobakiHC00}. 
Moreover, adapting a model to the child speech domain often degrades performance on adult speech due to domain mismatch~\cite{wang2026mind}. 
Even within child speech, variations across age groups and {recording environments introduce pronounced acoustic differences~\cite{ying25_wocci,shetty2025enhancing,shi2025comparing}}, making it challenging for a single model to robustly handle all domains simultaneously.



To address these gaps, we investigate the effectiveness of Speech-LLMs for child ASR and explore how a single Speech-LLM can robustly handle ASR across heterogeneous speech domains, encompassing both adult and child speech across diverse recording environments and age groups.

Specifically, we leverage the Mixture-of-Experts (MoE) paradigm~\cite{shazeer2017outrageously,fedus2022review} to enhance the ability of Speech-LLMs to handle heterogeneous speech domains, with each expert specializing in a specific domain.
Most recent MoE-based Speech-LLM approaches~\cite{CappellazzoKPFB25,lei2026moe,li2025mosa,pandey2026dynamic} typically rely on trainable gating networks without explicit guidance for expert routing. In contrast, our framework introduces a Classifier-based Domain Router (C-DR), enabling controllable and interpretable expert routing. Different speech domains exhibit varying levels of divergence, with large differences between adult and child speech or across distinct acoustic environments, whereas differences among other domains, such as neighboring child age groups (e.g., 4--7 vs.\ 8--10 years) within the same environment, can be relatively subtle~\cite{zheng25_wocci}. To reflect this hierarchy, we adopt a coarse-to-fine domain classification strategy for the C-DR. Moreover, routing uncertainty naturally arises during inference, particularly for utterances lying near domain boundaries (e.g., age ranges with overlapping acoustics). To address this, we introduce Entropy-Aware Routing (EAR), which dynamically incorporates a shared expert trained on data aggregated from multiple domains based on routing uncertainty. Beyond routing design, existing MoE-based Speech-LLM approaches typically introduce MoE either at the projector level~\cite{CappellazzoKPFB25,lei2026moe,li2025mosa,pandey2026dynamic} or at the LLM-LoRA level~\cite{MuWSXX25,mu2025mixture}, focusing primarily on either acoustic-domain or linguistic variability. In contrast, we combine a Mixture-of-Projectors (MoP) with a Mixture-of-LoRAs (MoL) to capture both domain-specific acoustic characteristics and LLM-side linguistic variations.

Experimental results demonstrate that our approach achieves strong child ASR performance while remaining robust across heterogeneous adult and child speech domains. Our main contributions are threefold: (1) we propose a unified Speech-LLM that integrates MoP and MoL to model both acoustic and linguistic variability; (2) we introduce a coarse-to-fine C-DR mechanism to capture hierarchical domain structures; and (3) we develop EAR to mitigate uncertain routing near domain boundaries. To our knowledge, this is the first Speech-LLM study to report strong results on public child ASR corpora while jointly analyzing adult, child, and age-specific performance.

\begin{figure*}[t]
    \centering
    \includegraphics[width=0.85\textwidth]{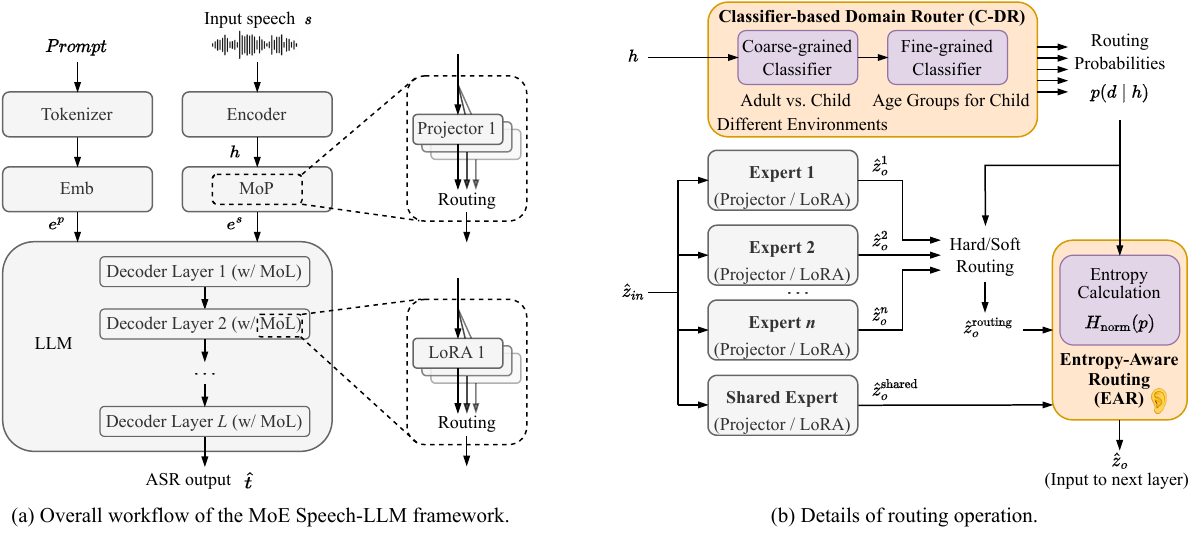}
    \vspace{-0.2cm}
    \caption{Illustration of the proposed method. (a) The overall workflow of the MoE Speech-LLM framework; dashed lines illustrate the mixture of projectors and the mixture of LoRAs. (b) Details of the routing operation, including the Classifier-based Domain Router and Entropy-Aware Routing. Each expert indicates either a domain-specific projector or LoRA. $\hat{z}_{in}$ and $\hat{z}_o^*$ denote the input and output of a projector or LoRA, respectively, unified for notational simplicity. The shared expert is trained on data aggregated from multiple domains.}
    \label{fig:0}
    \vspace{-0.8cm}
\end{figure*}
\vspace{-0.2cm}
\section{Background}
\vspace{-0.1cm}
We study ASR across heterogeneous speech domains, including adult and child speech from diverse acoustic environments and age groups, with the goal of building a unified multi-domain ASR system using a single Speech-LLM framework.

A typical Speech-LLM for ASR~\cite{FathullahWLJSLG24,ShiJXXZWSZY24} consists of a speech encoder, a modality projector, and an LLM fine-tuned with LoRA. Given a speech signal $s$, the transcription is given by:

\vspace{-0.4cm}

\begin{align}
&h = \text{Encoder}(s), \quad
e^s = \text{Projector}(h), \\
&e^p = \text{Emb}(\text{Tokenizer}(prompt)),\quad
\hat{t} = \text{LLM}(e^p, e^s),
\end{align}
where \texttt{Tokenizer} and \texttt{Emb} denote the tokenizer and text embedding layers of the LLM, and $\hat{t}$ is the predicted transcription.

LoRA~\cite{HuSWALWWC22} enables parameter-efficient adaptation of the LLM by injecting trainable low-rank updates into selected weight matrices while keeping the pre-trained parameters frozen. 
For a linear layer with weight matrix $W \in \mathbb{R}^{b \times a}$, LoRA re-parameterizes it as $W' = W + BA$, where $A \in \mathbb{R}^{r \times a}$ and $B \in \mathbb{R}^{b \times r}$ are trainable matrices with rank $r \ll \min(b,a)$. 
During fine-tuning, only $A$ and $B$ are optimized.

\vspace{-0.1cm}
\section{Method}
\vspace{-0.1cm}

\begin{table*}[t]
\centering
\caption{WER (\%) comparison of different methods across test domains. Bold font denotes the best results excluding upper-bound results. C-DR MoE with a weighted-layer classifier achieves the best results on MyST without affecting Libri-Clean. Under soft routing, applying EAR to OGI-S achieves the best results on OGI-S, while maintaining performance on MyST and Libri-Clean. For EAR, the shared expert (projector and LoRA) is taken from the model trained on the full OGI-S training set, corresponding to the second row of the upper-bound results. $^*$ indicates statistical significance ($p < 0.05$) compared to the best baseline, and $^\dagger$ indicates the best result from the Open ASR Leaderboard.}
\vspace{-0.25cm}
\label{tab:results}
\small 
\resizebox{0.9\textwidth}{!}{
\begin{tabular}{ll cccc cc}
\toprule
\multicolumn{2}{l}{\multirow{2}{*}{\textbf{Method}}} & \multicolumn{4}{c}{\textbf{OGI-S}} & \multirow{2}{*}{\textbf{MyST}} & \multirow{2}{*}{\textbf{Libri-Clean}} \\
\cmidrule(lr){3-6}
& & Age 4-7 & Age 8-10 & Age 11-15 & Avg. & & \\
\midrule

\multirow{3}{*}{Upper-bound} 
& Single Dataset Fine-tune (Encoder+Projector+LoRA) & 17.32 & 10.45 & 8.34 & 10.93 & 8.34 & - \\
& Single Dataset Fine-tune (Projector+LoRA)     & 19.15 & 10.57 & 9.29 & 11.83 & 8.58 & - \\
& Previous SOTA~\cite{FanSA24}~\cite{ying25_wocci}        & - & - & - & 11.6 & 8.5 & $\text{1.4}^\dagger$ \\

\midrule
\multirow{5}{*}{Baseline} 
& Zero-shot Canary-Qwen            & 24.97 & 14.86 & 13.34 & 16.31 & 8.96 & 1.61 \\
& Single-Expert        & 20.35 & 12.50 & 11.07 & 13.63 & 9.27 & 2.26 \\
& Vanilla-Routing MoE (Joint) & 20.30 & 11.55 & 10.37 & 12.89 & 8.74 & 2.37 \\
& Vanilla-Routing MoE (Pretrain + Joint) & 19.55 & 11.30 & 10.53 & 12.73 & 8.63 & 2.31 \\
& Vanilla-Routing MoE (Pretrain + Gate-only) & 20.02 & 10.81 & 9.26 & 12.08 & 12.40 & 1.96 \\

\midrule
\multirow{5}{*}{\shortstack[l]{C-DR MoE\\(Hard Routing)}} 
& Ground-Truth Routing      & 18.65 & 10.34 & \textbf{8.62}\rlap{$^*$} & 11.33 & \textbf{8.58} & \textbf{1.61} \\
& Top-layer Classifier (Single-Stage)      & 18.68 & 10.72 & 8.72 & 11.50 & 8.61 & 1.65 \\
& Weighted-layer Classifier (Single-Stage) & 18.52 & 10.48 & 8.70 & 11.39 & \textbf{8.58} & \textbf{1.61} \\
& Weighted-layer Classifier (Coarse-to-Fine)   & 18.43 & 10.41 & 8.66 & 11.32 & \textbf{8.58} & \textbf{1.61} \\
& \quad \quad + EAR for OGI-S        & 17.86 & 10.33 & 8.63 & 11.16 & \textbf{8.58} & \textbf{1.61} \\

\midrule
\multirow{2}{*}{\shortstack[l]{C-DR MoE\\(Soft Routing)}} 
& Weighted layer classifier (Coarse-to-Fine)      & 18.27 & 10.41 & 8.72 & 11.31 & \textbf{8.58} & \textbf{1.61}  \\
& \quad \quad + EAR for OGI-S & \textbf{17.64}\rlap{$^*$} & \textbf{10.28}\rlap{$^*$} & \textbf{8.62}\rlap{$^*$} & \textbf{11.08}\rlap{$^*$} & \textbf{8.58} & \textbf{1.61} \\

\bottomrule
\end{tabular}
}
\vspace{-0.4cm}
\end{table*}

\subsection{Overall Framework}
\vspace{-0.15cm}
Figure~\ref{fig:0} illustrates the overall workflow of the proposed framework. Unlike standard Speech-LLMs with a single projector and LoRA module, our framework adopts a Mixture-of-Experts (MoE) architecture to handle heterogeneous speech domains.
We introduce an explicit Classifier-based Domain Router (C-DR) to guide expert selection, enabling controllable and interpretable routing. In addition, the framework incorporates both a Mixture-of-Projectors (MoP) and a Mixture-of-LoRAs (MoL), where each domain is associated with a dedicated projector–LoRA pair. For simplicity, we refer to each domain-specific projector or LoRA as an \emph{expert}.

{During inference, the encoder representation $h$ is fed into the C-DR to produce routing probabilities $p(d \mid h)$,
where $d \in \{1,\dots,n\}$ indexes the domain. These probabilities are used to aggregate expert outputs via a routing operation. Entropy-Aware Routing (EAR) is further applied during routing to address the issue of routing uncertainty. Details of the C-DR, routing and EAR are described in Sections~\ref{sec:router} and~\ref{sec:entropy_routing}.}



\vspace{-0.25cm}
\subsection{Domain Router Based on Coarse-to-Fine Classification}
\label{sec:router}
\vspace{-0.15cm}
Speech domains often exhibit a hierarchical structure, with large differences across coarse-grained domains (e.g., adult vs.\ child speech or distinct acoustic environments) and more subtle differences across fine-grained domains such as child age groups. To model this hierarchy, we employ a coarse-to-fine classification strategy for the C-DR.

A coarse-grained classifier first predicts high-level domains. For inputs assigned to coarse domains with finer distinctions, a corresponding fine-grained classifier produces routing logits for sub-domains. The final routing probabilities $p(d \mid h)$ are obtained by combining the coarse and fine classifier outputs. Both classifiers are linear models operating on a weighted sum of multi-layer encoder representations, with learnable layer weights, followed by softmax normalization.

Given $p(d \mid h)$ and the output of each expert $\hat{z}_{o}^{d}$, expert selection is performed via either hard or soft routing. Hard routing selects the most probable expert,
\vspace{-0.1cm}
\begin{align}
d^{*} = \arg\max_{d} p(d \mid h), \quad
\hat{z}_o^{\text{routing}} = \hat{z}_{o}^{d^{*}},
\end{align}
whereas soft routing aggregates expert outputs using probability-weighted averaging,
\vspace{-0.1cm}

\begin{equation}
\hat{z}_o^{\text{routing}} = \sum_{d=1}^{n} p(d \mid h)\, \hat{z}_{o}^{d}.
\end{equation}

\vspace{-0.1cm}
\subsection{Entropy-Aware Routing}
\label{sec:entropy_routing}
\vspace{-0.1cm}

During inference, routing probabilities from the C-DR can become uncertain when processing acoustically ambiguous inputs, {particularly for utterances lying near domain boundaries (e.g., age ranges with overlapping acoustics).} To address this issue, we propose Entropy-Aware Routing (EAR), which leverages routing uncertainty to dynamically incorporate a \emph{shared expert} trained on data aggregated across domains.

Given routing probabilities $p(d \mid h)$, we quantify routing uncertainty using the normalized entropy $H_{\text{norm}}(p)\in[0,1]$:
\begin{equation}
H_{\text{norm}}(p) = -\frac{1}{\log n} \sum_{d=1}^{n} p(d \mid h) \log p(d \mid h),
\end{equation}
where higher $H_{\text{norm}}(p)$ indicates lower routing confidence.

EAR interpolates between domain-specific and shared experts to produce the output of the MoE layer based on $H_{\text{norm}}(p)$:
\begin{equation}
\hat{z}_o = (1 - H_{\text{norm}}(p)) \, \hat{z}_o^{\text{routing}} + H_{\text{norm}}(p) \, \hat{z}_o^{\text{shared}}.
\end{equation}
where $\hat{z}_o^{\text{routing}}$ is obtained via standard routing (Section~\ref{sec:router}). This enables smooth expert blending under routing uncertainty.




\vspace{-0.15cm}
\subsection{Training and Inference}
\vspace{-0.1cm}
During training, the MoE Speech-LLM and the C-DR are optimized separately. For the MoE Speech-LLM, the encoder and LLM backbone are frozen, and only domain-specific projectors and LoRA modules (experts) are trained using ground-truth hard routing. {For the shared expert, a separate set of projector and LoRA parameters is trained within the same frozen Speech-LLM backbone using data aggregated from multiple domains.} For C-DR training, the frozen encoder is reused, and two learnable weighted combinations of multi-layer encoder representations are applied for coarse- and fine-grained classification.

During inference, the C-DR is integrated into the MoE Speech-LLM to guide expert routing, following the strategies described in Sections~\ref{sec:router} and~\ref{sec:entropy_routing}.

\vspace{-0.3cm}
\section{Experimental Settings}
\vspace{-0.1cm}



\subsection{Dataset Setting}
\vspace{-0.2cm}
We evaluate the proposed method on both child and adult speech. For child speech, we use MyST~\cite{PradhanCW24} and the spontaneous portion of OGI (OGI-S)~\cite{ShobakiHC00}, which are collected in different acoustic environments and include disfluency-preserving transcriptions. MyST contains dialogues between children aged 8--10 and virtual tutors, while OGI-S consists of classroom-recorded responses from a broader age range. Following~\cite{ying25_wocci}, both datasets are preprocessed, and OGI-S is further divided into three age groups: 4--7, 8--10, and 11--15. For adult speech, we adopt the LibriSpeech~\cite{PanayotovCPK15} test-clean subset (Libri-Clean) for evaluation.
Different datasets are treated as coarse-grained domains based on differences between child and adult speech and recording environments (e.g., MyST vs.\ OGI-S), while age groups within OGI-S form fine-grained domains capturing age-dependent variation. This yields five domains in total.

\vspace{-0.25cm}
\begin{figure}[t]
    \centering
    \includegraphics[width=0.9\columnwidth]{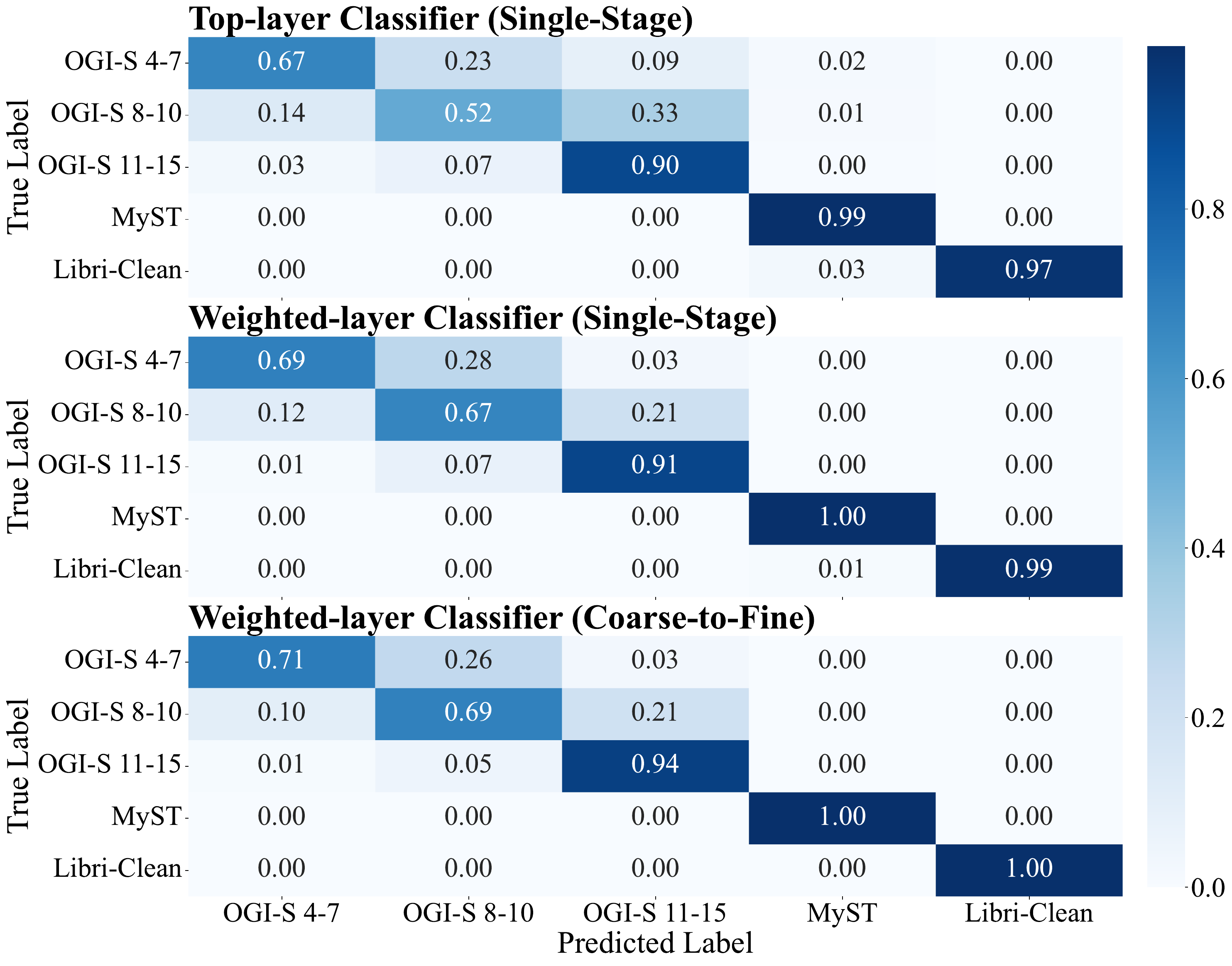}
    \vspace{-0.2cm}
    \caption{Confusion matrices of domain classification under 3 settings: (1) top-layer encoder representation (single-stage classification), (2) weighted-sum encoder representation (single-stage classification), and (3) weighted-sum encoder representation (coarse-to-fine classification).}
    \label{fig:1}
    \vspace{-0.2cm}
\end{figure}
\vspace{-0.05cm}
\subsection{Training and Inference Setting}
\vspace{-0.15cm}
We adopt Canary-Qwen~\cite{nvidia_canary_qwen_25b}, a top-performing model on Open ASR Leaderboard~\cite{open_asr_leaderboard}, as the pre-trained Speech-LLM backbone.
The C-DR is trained on the training sets of MyST, OGI-S, and LibriSpeech train-clean-100 using the proposed coarse-to-fine classification strategy with weighted-sum encoder embeddings. 
We additionally evaluate alternative router designs based on top-layer embeddings and single-stage classification.

For MoE Speech-LLM fine-tuning, the projector and LoRA parameters of Canary-Qwen are replicated to initialize five experts. The LoRA configuration follows the implementation of Canary-Qwen.
Experts corresponding to MyST and the three OGI-S age groups are fine-tuned with ground-truth hard routing, while the adult expert is kept fixed.
All models are trained on a single NVIDIA A6000 GPU with a batch size of 2 and gradient accumulation of 8.
We use the AdamW optimizer with a peak learning rate of $1\times10^{-4}$ and a cosine warmup–decay schedule with 500 warmup steps and 60{,}000 total steps.

{Since fine-grained differences among age groups within OGI-S are substantially smaller than the coarse-grained differences across datasets, and domain classification across different datasets is highly accurate in our experiments, we use age-group distinctions within OGI-S as a representative setting to evaluate the proposed EAR. Accordingly, EAR is applied only to OGI-S age groups, with entropy computed from probabilities over the three age groups.}
The shared expert is trained on the full OGI-S training set.
ASR is evaluated using word error rate (WER).

\vspace{-0.3cm}
\subsection{Baselines and Upper Bounds}
\label{sec:baseline}
\vspace{-0.15cm}

\subsubsection{Main Baselines.}
\vspace{-0.15cm}
The baselines include:
(1) a \emph{zero-shot} Canary-Qwen evaluated without any fine-tuning;
(2) a \emph{single-expert} Canary-Qwen with one projector and one LoRA module, jointly fine-tuned on all datasets while freezing the encoder and the LLM backbone; and
(3) a \emph{vanilla-routing MoE} with five projectors and five LoRA modules, where utterance-level expert selection is governed by trainable gating networks~\cite{lei2026moe} rather than an explicit classifier.

For the vanilla-routing MoE baseline, we consider three training strategies:
{\emph{(Joint)}, where experts and routing gates are trained jointly; \emph{(Pretrain + Joint)}, where experts are first pre-trained with ground-truth hard routing and then jointly trained with routing gates; and \emph{(Pretrain + Gate-only)}, where experts are pre-trained and only routing gates are trained afterward.}

\vspace{-0.25cm}
\subsubsection{Upper Bounds.}
\vspace{-0.15cm}
To estimate dataset-specific upper bounds, we report results from (1) single-dataset fine-tuning of the Speech-LLM on the MyST and OGI-S training sets under two settings, where either only the projector and LoRA are updated or the encoder, projector, and LoRA are jointly updated, and (2) previously published state-of-the-art (SOTA) results. We do not include fine-tuning for Libri-Clean, as its zero-shot performance is already strong.

\begin{figure}[t]
    \centering
    \includegraphics[width=0.9\columnwidth]{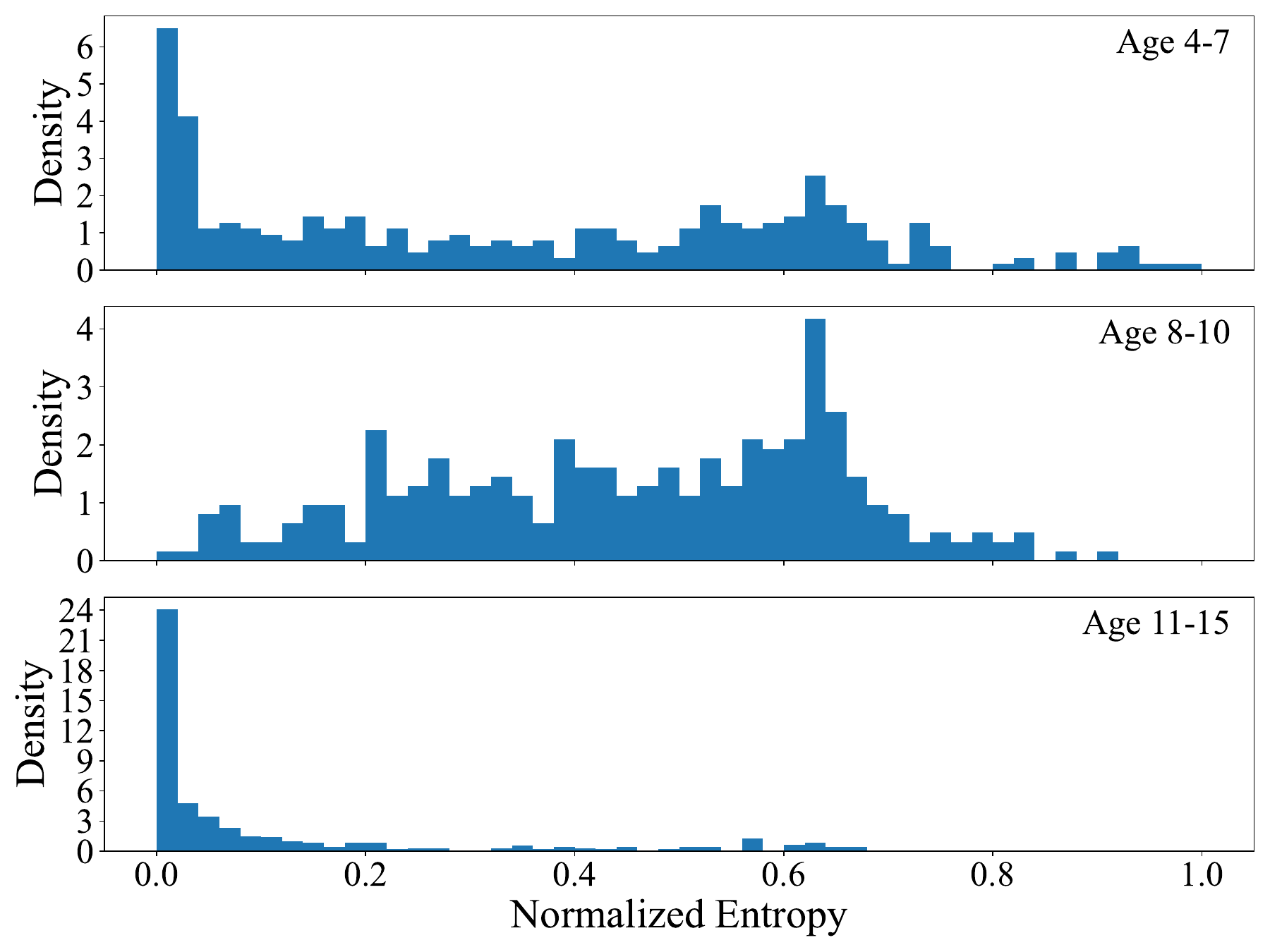}
    \vspace{-0.3cm}
    \caption{Distribution of normalized entropy of fine-grained routing probabilities for the three age groups in OGI-S.}
    \label{fig:2}
    \vspace{-0.25cm}
\end{figure}
\vspace{-0.3cm}
\section{Experimental Results}
\vspace{-0.1cm}
\subsection{Performance of Domain Classification}
\vspace{-0.2cm}
Figure~\ref{fig:1} shows confusion matrices for three domain classifier variants:
top-layer encoder representations,
weighted-sum representations with single-stage classification,
and weighted-sum representations with coarse-to-fine classification.
Classification across different datasets is fairly accurate.
In contrast, age-group classification within OGI-S remains more challenging.
Compared to top-layer features, weighted-sum representations consistently improve accuracy by leveraging information from intermediate encoder layers.
The proposed coarse-to-fine strategy further improves performance on OGI-S, as the fine-grained classifier better captures subtle age-dependent variations.

\vspace{-0.35cm}
\subsection{ASR Performance across Test Domains}
\vspace{-0.15cm}
Table~\ref{tab:results} compares WER across test domains for upper-bound systems, baselines, and the proposed C-DR guided MoE.
Upper-bound results show that dataset-specific fine-tuning of Canary-Qwen achieves new SOTA on both OGI-S and MyST, highlighting the effectiveness and potential of Speech-LLMs for child ASR.
In contrast, the single-expert baseline performs poorly due to its inability to balance heterogeneous domains, with training on OGI-S even degrading performance on MyST and Libri-Clean relative to zero-shot inference.

{The vanilla-routing MoE with 3 different training settings yields improvements in some cases but fails to achieve consistent gains across domains, likely due to the lack of explicit domain supervision and limited data for training the router.} In comparison, the proposed C-DR guided MoE consistently outperforms all baselines.
Benefiting from more accurate domain classification, the weighted-layer classifier with a coarse-to-fine strategy generally outperforms both top-layer and single-stage variants and achieves the best results on MyST (excluding the upper bound) without affecting Libri-Clean. Under soft routing, applying EAR to OGI-S further yields the best results on OGI-S (excluding the upper bound). Notably, although the shared expert alone attains a relatively poor average WER of 11.83\% (19.15\%, 10.57\%, and 9.29\% for the three age groups), EAR still yields additional gains, demonstrating the effectiveness of uncertainty-aware expert incorporation. Overall, combining soft routing with EAR yields the best performance on OGI-S.
Figure~\ref{fig:2} visualizes the normalized routing entropy on the OGI-S test set. Routing decisions are most certain for the 11--15 age group, while the other age groups exhibit higher uncertainty. {This behavior may be attributed to the larger acoustic gap between younger and older children, making expert assignment less confident for younger age groups.}

{Furthermore, we observe that for the 4--7 age group in OGI-S, routing based on classifier-predicted probabilities can even outperform routing based on ground-truth age labels, with EAR yielding the most pronounced improvement in this group. This behavior may be attributed to the fact that pronunciation characteristics do not always strictly align with chronological age, as children of the same age can be undergoing different stages of speech development. These observations further emphasize the effectiveness and necessity of EAR for child age groups, as it leverages expert ensembles and uncertainty-based interpolation to better handle acoustically ambiguous inputs.}

\vspace{-0.35cm}
\subsection{Ablation on Model Structure}
\vspace{-0.2cm}
Our MoE Speech-LLM incorporates both a Mixture of Projectors (MoP) and a Mixture of LoRAs (MoL). In Table~\ref{tab:ablation_moe}, we compare our full model with two ablated variants: MoP only (using a single LoRA) and MoL only (using a single projector), both evaluated under ground-truth routing.
The results show that removing either component leads to reduced performance, demonstrating the complementary contributions of MoP and MoL. Specifically, MoP-only outperforms MoL-only, suggesting that modeling domain-specific acoustic variability via projectors is more critical than modeling linguistic variations via LoRAs applied to the LLM. Although using multiple projectors and LoRAs increases the number of parameters, this overhead is modest (about 5\%) relative to the Speech-LLM backbone.

\begin{table}[t]
\centering
\caption{Ablation study on model structure comparing models with and without a mixture of projectors (MoP) and a mixture of LoRAs (MoL), evaluated under ground-truth hard routing. Bold font denotes the best results, $^*$ indicates statistical significance with $p <0.05$.}
\vspace{-0.25cm}
\label{tab:ablation_moe}
\resizebox{\columnwidth}{!}{
\begin{tabular}{@{} cc ccc cc @{}} 
\toprule
\multirow{2}{*}{\textbf{MoP}} & \multirow{2}{*}{\textbf{MoL}} & \multicolumn{3}{c}{\textbf{OGI-S}} & \multirow{2}{*}{\textbf{MyST}} & \multirow{2}{*}{\textbf{Libri-Clean}} \\
\cmidrule(lr){3-5} 
& & Age 4--7 & Age 8--10 & Age 11--15 & & \\
\midrule
\cmark & \xmark & 18.94 & 10.53 & 9.21 & 8.81 & 2.13 \\
\xmark & \cmark & 20.91 & 11.59 & 9.97 & 8.96 & 2.22 \\
\cmark & \cmark & \textbf{18.65}\rlap{$^*$} & \textbf{10.34}\rlap{$^*$} & \textbf{8.62}\rlap{$^*$} & \textbf{8.58}\rlap{$^*$} & \textbf{1.61}\rlap{$^*$} \\
\bottomrule
\end{tabular}
}
\end{table}

\vspace{-0.3cm}
\section{Conclusion}
\vspace{-0.2cm}


We propose an Entropy-Aware Domain-Routed MoE Speech-LLM to tackle ASR across heterogeneous speech domains, including adult and child speech with diverse conditions. By combining mixtures of projectors and LoRA modules, along with coarse-to-fine domain classification and entropy-aware routing, our model effectively captures domain-specific variations and handles age-related subtleties. Experiments show that it consistently outperforms single-expert baselines and vanilla MoE models across multiple test domains. To our knowledge, this is the first Speech-LLM framework capable of simultaneously handling multiple child and adult speech domains, and future work will explore generalization to additional domains.


\newpage
\section{Acknowledgements}
\label{sec:ack}
This research is supported in part by the National Science Foundation (NSF) and the Institute of Education Sciences (IES), U.S. Department of Education (DoE), through Grant R305C240046 to the U. at Buffalo. The opinions expressed are those of the authors and do not represent views of the IES, DoE, or the NSF. 
\section{Generative AI Use Disclosure}
\label{sec:genai_disclosure}
During the preparation of this work, the authors used ChatGPT (GPT-5.2 Thinking) for language editing, including proofreading and improving clarity and readability of the manuscript. All technical content, experimental design, results, and conclusions were produced and verified by the authors. After the use of Generative AI, the authors reviewed and edited the manuscript and take full responsibility for the content of the publication. Generative AI tools were not used to produce a significant portion of the manuscript and are not listed as authors.


\end{document}